\documentclass{moriond}

\def\be{\begin{equation}}
\def\ee{\end{equation}}
\def\bea{\begin{eqnarray}}
\def\eea{\end{eqnarray}}

\usepackage{bm,amsmath,amssymb,units}

\newcommand{\Photo}{\vspace{0.2cm} \includegraphics[height=35mm]{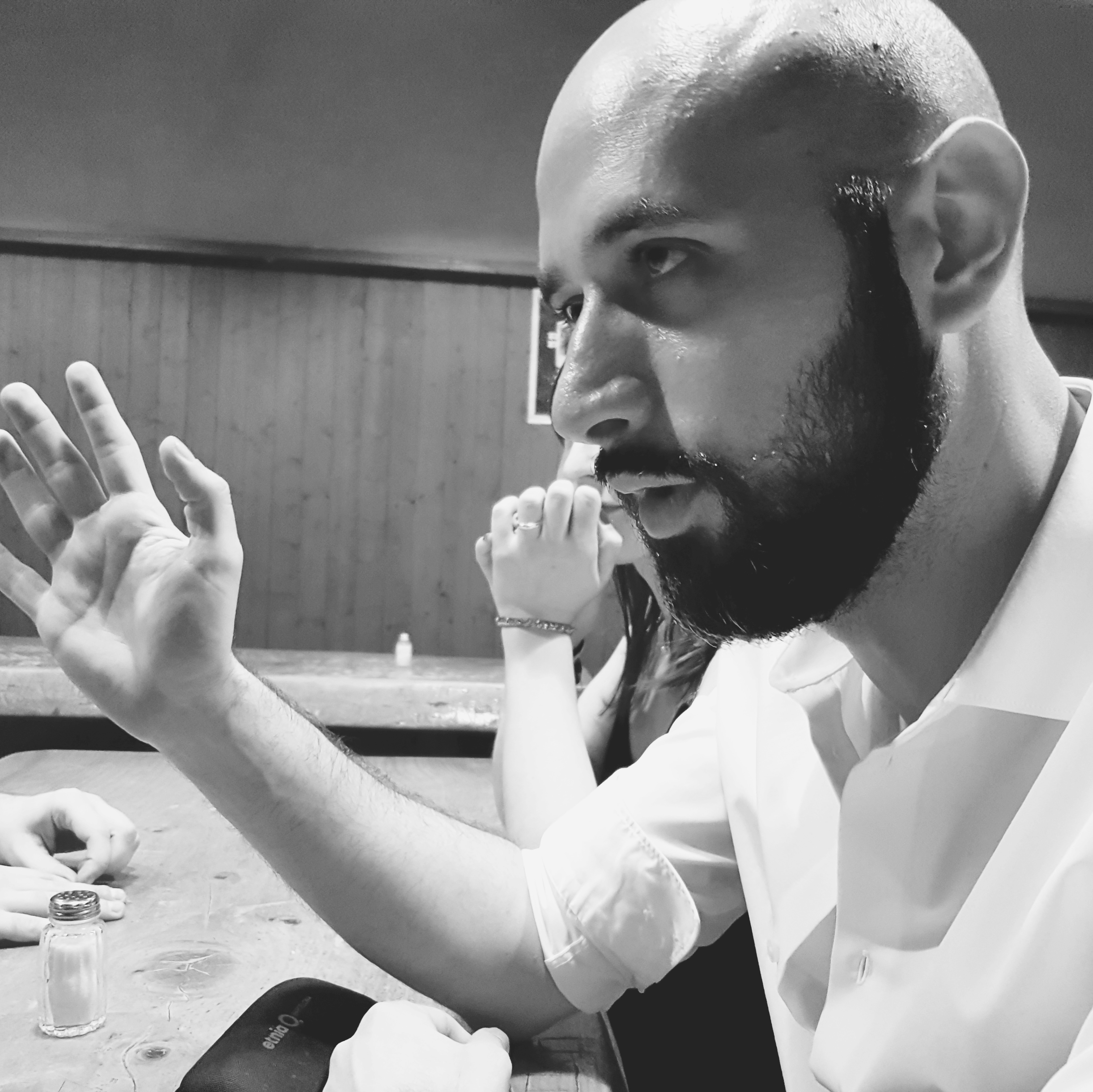}}

\begin{document}
\begin{flushright}
ULB-TH/17-10
\end{flushright}
\vspace*{4cm}
\title{CLOCKWORK DARK MATTER}

\author{ D. TERESI }

\address{Service de Physique Th\'eorique, Universit\'e Libre de Bruxelles, Boulevard du Triomphe, CP225, 1050 Brussels, Belgium}

\maketitle\abstracts{
I give a pedagogical discussion of thermal dark matter (DM) within the clockwork mechanism. The clockwork mechanism, which is a natural way to generate small numbers starting from order-one couplings, allows to have a long-lived, but unstable, DM particle that nevertheless has $O(1)$ couplings with electroweak- or TeV-scale states. Remarkably, DM decays on time scales much longer than the age of the Universe and has at the same time sizeable couplings with light states, which therefore allow to produce it thermally within the WIMP paradigm.
These new particles with large couplings can be searched for at current or future colliders. I also briefly comment on how this setup can minimally emerge from the deconstruction of an extra dimension in \emph{flat} spacetime.}

\section{Introduction}
The clockwork mechanism~\cite{Choi:2015fiu,Kaplan:2015fuy,Giudice:2016yja} is an elegant and economical way to generate tiny couplings much smaller than unity, or equivalently large hierarchies between different dimensionful or dimensionless quantities, in a theory with only $O(1)$ couplings and hierarchies. The important point here is that the mechanism is \emph{economical}, in the sense that to generate a hierarchy $X$ one only needs a number of fields $O(\log X)$. This should be contrasted to what happens in different setups, e.g. the one of Ref.~\cite{Arkani-Hamed:2016rle}, where  a number of fields $O(X)$ are needed.

The mechanism was originally introduced in the context of relaxion models~\cite{Choi:2015fiu,Kaplan:2015fuy}, to overcome a theoretical difficulty present generically there, namely the fact that the relaxion field needs to have a huge super-Planckian excursion during the relaxation phenomenon, together with the fact that tiny (with respect to the other scales involved) dimensionful couplings are needed in the original formulation of the relaxion mechanism~\cite{Graham:2015cka}. 

However, it has been recently shown that the clockwork mechanism is much more general than that~\cite{Giudice:2016yja}, and that it could be indeed useful for a number of physical situations where large hierarchies/small couplings are indeed present~\cite{Giudice:2016yja,Kehagias:2016kzt,Farina:2016tgd,Ahmed:2016viu,Hambye:2016qkf}: axion models that are ``invisible'' although the relevant physics is at the weak scale~\cite{Giudice:2016yja,Farina:2016tgd}, cosmic inflation~\cite{Kehagias:2016kzt}, the flavour puzzle and, most remarkably, the long-standing (and these days more disturbing than ever) gauge hierarchy problem~\cite{Giudice:2016yja}.
Last but not least, dark matter~\cite{Hambye:2016qkf}, which is the main focus of this work and that I will now review, following the original study of Ref.~\cite{Hambye:2016qkf}.

Dark matter (DM) must be stable over cosmological time scales. This means that it could be absolutely stable or very long lived.  Absolute stability is typically realized by invoking a symmetry, either continuous or discrete, local or global, motivated or \emph{ad-hoc}. This is the most popular option and is certainly fine. On the other hand, the possibility that DM may be unstable, but long lived, is in turn very interesting as its decay could be probed through indirect-detection searches. In this case the lifetime of a DM candidate in the WIMP mass range, which I will consider in the sequel, must be typically larger than $\tau \sim 10^{26}$~sec. To obtain such a large lifetime, one may typically assume that the decay is induced by the exchange of very heavy particles. From the effective field theory perspective at low energy, assuming couplings of order one, instability associated to a dimension-5 operator would require that the heavy degrees of freedom have masses above the Planck scale. The situation is better for dimension-6 operators, although still these particles need to be around the GUT scale or heavier. In alternative, one could assume that the particles that trigger the decay of DM are much lighter, but then with tiny coupling. In all these cases, the physics involved in the decay of DM would be essentially impossible to test.

The clockwork mechanism can come to our aid. In Ref.~\cite{Hambye:2016qkf} we have shown that,  by making use of the clockwork mechanism, a DM particle could be made very long lived, at these same time having a decay into SM particles induced by $O(1)$ interactions with particles that could be produced at colliders, and therefore tested directly and indirectly. Such  $O(1)$ interactions make DM annihilate into SM or hidden-sector particles fast, so that its relic density is determined by the standard freeze-out mechanism. We arrive at the (at first sight paradoxical) fact that DM is unstable and at the same a WIMP. 

One may still think that $O(\log X)$ fields are too many, and that the clockwork mechanism is not that economical, after all. For the appreciators of minimalism the good news is that the clockwork setup may originate from the deconstruction of an extra dimension, in which only one or few fields are introduced. Although in the original Ref.~\cite{Giudice:2016yja} such a construction involved a curved spacetime (with an exponential metric in the extra dimension), we have subsequently shown~\cite{Hambye:2016qkf} that the clockwork setup can originate from a \emph{flat} extra dimension and boundary terms, as I will briefly review below. It has been subsequently claimed that this is the only way to obtain consistently the clockwork mechanism from an extra dimension~\cite{Craig:2017cda}, although very recently it has been argued~\cite{Giudice:2017suc} that the curved- and flat-spacetime constructions are instead equivalent, at least from the low-energy perspective.

\section{Clockwork dark matter}
The clockwork mechanism is based on the very simple observation that a product such as:
\begin{equation}
1/q \quad \times \quad 1/q \quad \times \quad  1/q \quad \times \quad 1/q \quad \times \quad  \ldots \quad \times \quad 1/q \;,
\end{equation}
with $q>1$, quickly becomes tiny when the number of factors increases. To implement this idea in quantum field theory, one wants to introduce $N$ fields $\phi_i$, which interact schematically as 
\begin{equation}
{ \phi_0} \;\; \overset{1/q}{\rule{2em}{0.5pt}} \;\; \phi_1 \;\; \overset{1/q}{\rule{2em}{0.5pt}} \;\;\phi_2 \;\; \overset{1/q}{\rule{2em}{0.5pt}} \;\;\phi_3 \;\; \overset{1/q}{\rule{2em}{0.5pt}} \;\; \ldots  \;\; \overset{1/q}{\rule{2em}{0.5pt}} \;\;\phi_N \;\; \overset{}{\rule{2em}{0.5pt}} \;\;{SM} \;,
\end{equation}
with couplings $1/q \lesssim 1$. The key feature of the clockwork mechanism is that, if thanks to an appropriate symmetry one of the mass eigenstates $\phi_{\it light}$ (typically the lightest one) is essentially given by $\phi_0$, its interaction with the Standard Model (SM) will be suppressed exponentially as
\begin{equation}
{ \phi_{\it light}\;
\raisebox{2pt}{\rule{1em}{0.5pt}} \; SM \sim 1/q^{N}} \;.
\end{equation}
Therefore, to generate a tiny coupling $1/X$ by making use of $O(1)$ couplings $1/q$, one only need a logarithmically large number of fields $N \sim \log_q X$. In the case of fermions, the convenient symmetry just mentioned can be taken to be the chiral symmetry. Thus, we introduce a set of chiral fermions $R_i$, $L_i$ interacting schematically as
\begin{equation} \label{eq:clockwork_scheme}
R_0 \;\; \overset{m}{\rule{1.5em}{0.5pt}} \;\; \underbrace{L_1 \quad  R_1}_{q m}  \;\; \overset{m}{\rule{1.5em}{0.5pt}} \;\; \underbrace{L_2 \quad  R_2}_{q m} \;\; \overset{m}{\rule{1.5em}{0.5pt}} \;\; \underbrace{L_3 \quad  R_3}_{q m}  \;\; \overset{m}{\rule{1.5em}{0.5pt}}  \; \ldots \;  \overset{m}{\rule{1.5em}{0.5pt}} \;\; \underbrace{L_N \quad  R_N}_{q m} \; \; \overset{}{\rule{1.5em}{0.5pt}} \;\; L_{SM} \;,
\end{equation}
where $L_{SM}$ is the SM lepton doublet. Since there are $2 N +1$ chiral symmetries
\begin{equation}
U(1)_{R_0} \times U(1)_{L_1} \times U(1)_{R_1} \times \ldots\times U(1)_{L_{N}} \times U(1)_{R_{N}} \;, \qquad
\text{with}  \quad U(1)_{R_{N}} \equiv U(1)_{L_{SM}} \;,
\end{equation}
and $2 N$ breaking parameters $m$, $q m$, there is a massless mode $N$. Intuitively, for $q>1$, since $R_0$ does not have a chiral partner, the massless mode $N$ approximately coincides with $R_0$. The conclusion is not changed if we add a Majorana mass $m_N < q m$, so that the light mode $N$ acquires a mass $\approx m_N$ and will be our dark matter particle. In order to realize the structure \eqref{eq:clockwork_scheme} we may consider a set of scalars $S_i$, $C_i$ with charges
\begin{equation}
S_i \sim (-1,1) \quad \text{under} \quad \text U(1)_{R_i} \times U(1)_{L_{i+1}} \; ,\qquad C_i \sim (1,-1) \quad \text{under} \quad \text U(1)_{L_i} \times U(1)_{R_{i}} \;.
\end{equation}
Thus, the fermion and scalar fields interact as a chain:
\begin{equation}\label{eq:scheme_scalars}
 R_0 \;\; \overset{S_1}{\rule{2em}{0.5pt}}  \;\; L_1 \;\; \overset{C_1}{\rule{2em}{0.5pt}} \;\;R_1 \;\; \overset{S_2}{\rule{2em}{0.5pt}} \;\;L_2\;\; \overset{C_2}{\rule{2em}{0.5pt}} \;\; \ldots  \;\; \overset{C_N}{\rule{2em}{0.5pt}} \;\;R_N \;\; \overset{}{\rule{2em}{0.5pt}} \;\;L_{SM} \;.
\end{equation}
When the scalars acquire vacuum expectation values we obtain the clockwork setup \eqref{eq:clockwork_scheme}, with $m= y_S \langle S_i \rangle$, $q m= y_C \langle C_i \rangle$, where $y_{S,C}$ denote the appropriate Yukawa couplings with the clockwork states.

\begin{figure}
\centering
\includegraphics[height=0.3\textwidth]{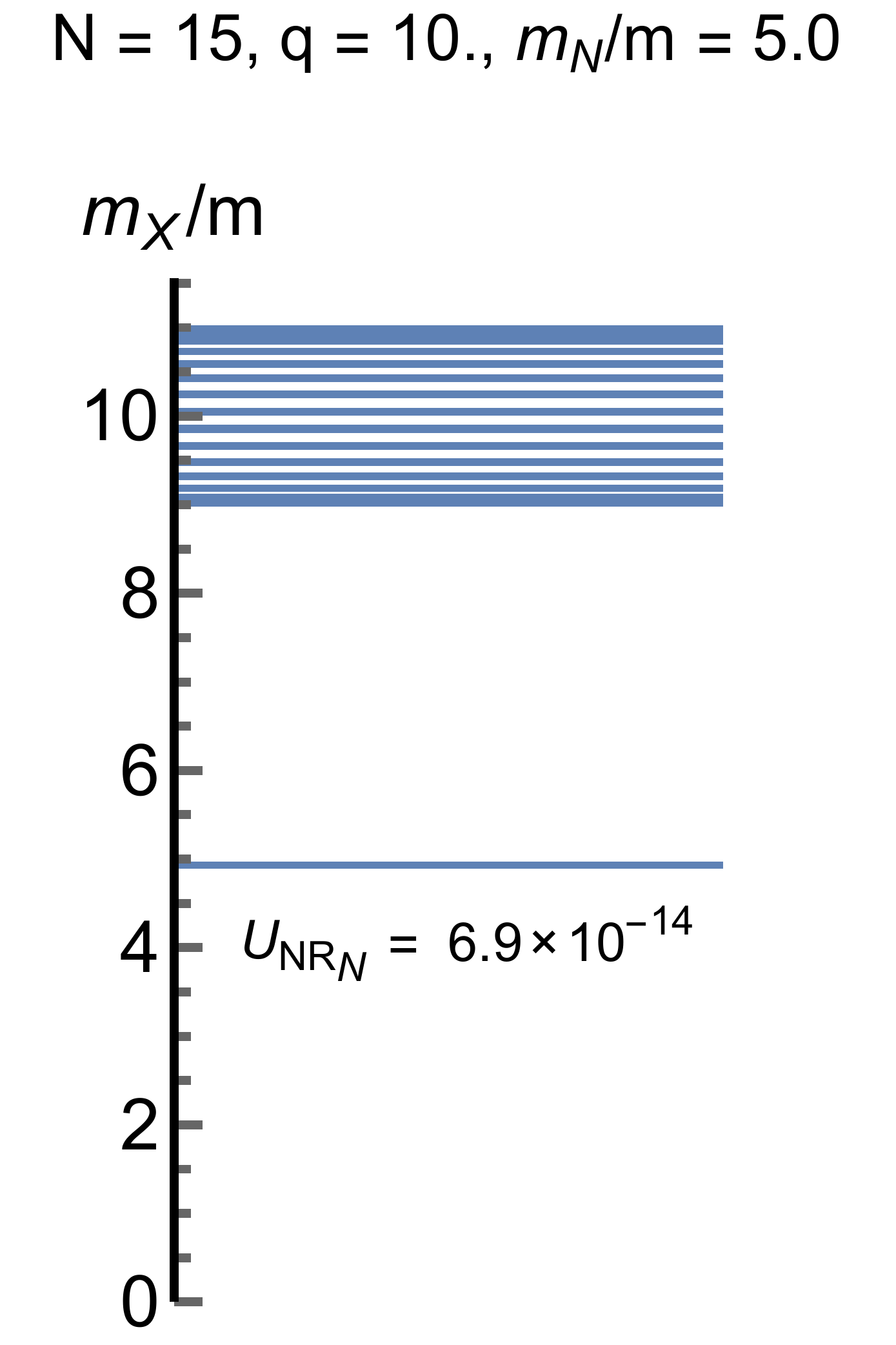}
\caption{Example spectrum for the clockwork dark matter framework, consisting of a light dark-matter particle $N$ and a band of clockwork gears $\psi_i$.\label{fig:spectrum}}
\end{figure}

From now on, let us take $q \gg 1$ for simplicity, so that an analytic understanding is facilitated. In this limit, one can show that the clockwork mechanism works as long as $m_N \lessapprox q m$. The fermionic spectrum of the theory consists of a the dark-matter Majorana fermion
\begin{equation}\label{eq:overlap}
N \;\approx \, R_0 \;+\; \frac{1}{q^1} R_1 \;+\; \frac{1}{q^2} R_2 \;+ \; \ldots \;+\; \frac{1}{q^N} R_N \;,
\end{equation}
and a band of $N$ Majorana pairs (the clockwork ``gears''), that form pseudo-Dirac states $\psi_i$ with mass $\approx q m$:
\begin{equation}
\psi_i \approx \frac{1}{\sqrt{N}} \sum_k \big( O(1) L_k \,+\, O(1) R_k  \big) \;.
\end{equation}
An example of the spectrum is shown in Fig.~\ref{fig:spectrum}. In addition, two sets of $N$ scalars $S_i$ and $C_i$ are  expected in the same mass range. However, these are not necessarily dynamic fields, for more details see Ref.~\cite{Hambye:2016qkf}.

The clockwork gears have sizeable interactions with the SM and the clockwork scalars. The dark-matter particle has sizeable interactions with the scalars too. For $q\gg 1$, the relevant $O(1)$ interactions are:
\begin{equation*}{\quad \parbox[c]{3cm}{\includegraphics[scale=0.3]{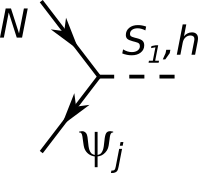}} \quad \parbox[c]{2.8cm}{\includegraphics[scale=0.3]{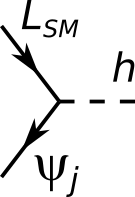}} \quad \parbox[c]{2cm}{\includegraphics[scale=0.3]{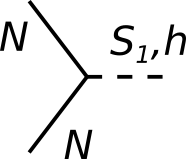}}} \;,
\end{equation*}
where $h$ is the SM Brout-Englert-Higgs scalar boson, which in the spirit of the clockwork mechanism (i.e.~that one does not have small interactions to start with in the Lagrangian) is expected to mix with the clockwork scalars, in particular $S_1$. Instead, the key aspect of the mechanism for clockwork dark matter is in the interactions of $N$ with the SM fermions. From \eqref{eq:overlap} we see that the overlap of $N$ with the state $R_i$ is approximately $1/q^i$. This is precisely because at each interaction with the clockwork pairs $L_i R_i$ along the chain, one gets a O(1) suppression $m/(qm) = 1/q$ (see \eqref{eq:clockwork_scheme}). Thus, if the SM leptons interact with the state $R_N$ only (with a O(1) Yukawa coupling $y$), in the original clockwork basis, the interaction of the mass eigenstate $N$ with them is exponentially clockwork suppressed
\begin{equation}
\mathcal L \supset -\, \frac{y}{q^N} \, \bar{L}_{SM} \widetilde{H} N_R \;,
\end{equation}
so that, even though $N$ can decay, e.g. $N \to \nu h , \nu Z , l W$, one could make \emph{the decay lifetime of $N$ longer than the age of the Universe by making use of only $\mathcal{O}(1)$ couplings and $\lesssim$ {\rm TeV}-scale states}. Indirect detection requires the lifetime to be typically larger than $\approx 10^{26}$ sec, i.e. $q^{2N} > 1.5 \times 10^{50} \,  y^2 \, m_N/\text{GeV}$. For instance, this can be realized with $m_N \sim 100 \, {\rm GeV}$, $y \sim 1$, $q \sim 10$, $ N \sim 26$. An important point to stress here is that one can show that potentially dangerous loop corrections involving the clockwork gears $\psi_i$, which have O(1) couplings and could potentially destabilize dark matter by generating large corrections to the decay processes, are also clockwork suppressed. This can be understood by noticing that any decay diagram for $N$ must involve a fermion line starting from the initial state, i.e.~$N \sim R_0$ itself, and ending into a (lighter) SM fermion in the final state: this can happen only if the fermion line proceeds along the whole clockwork chain in~\eqref{eq:clockwork_scheme}, thus inheriting the exponential clockwork suppression.

\section{Phenomenology}
Let us now see how this setup can be used to generate dark matter in the early Universe and the phenomenological consequences of this framework. 

\begin{figure}
\centering
\includegraphics[width=7cm]{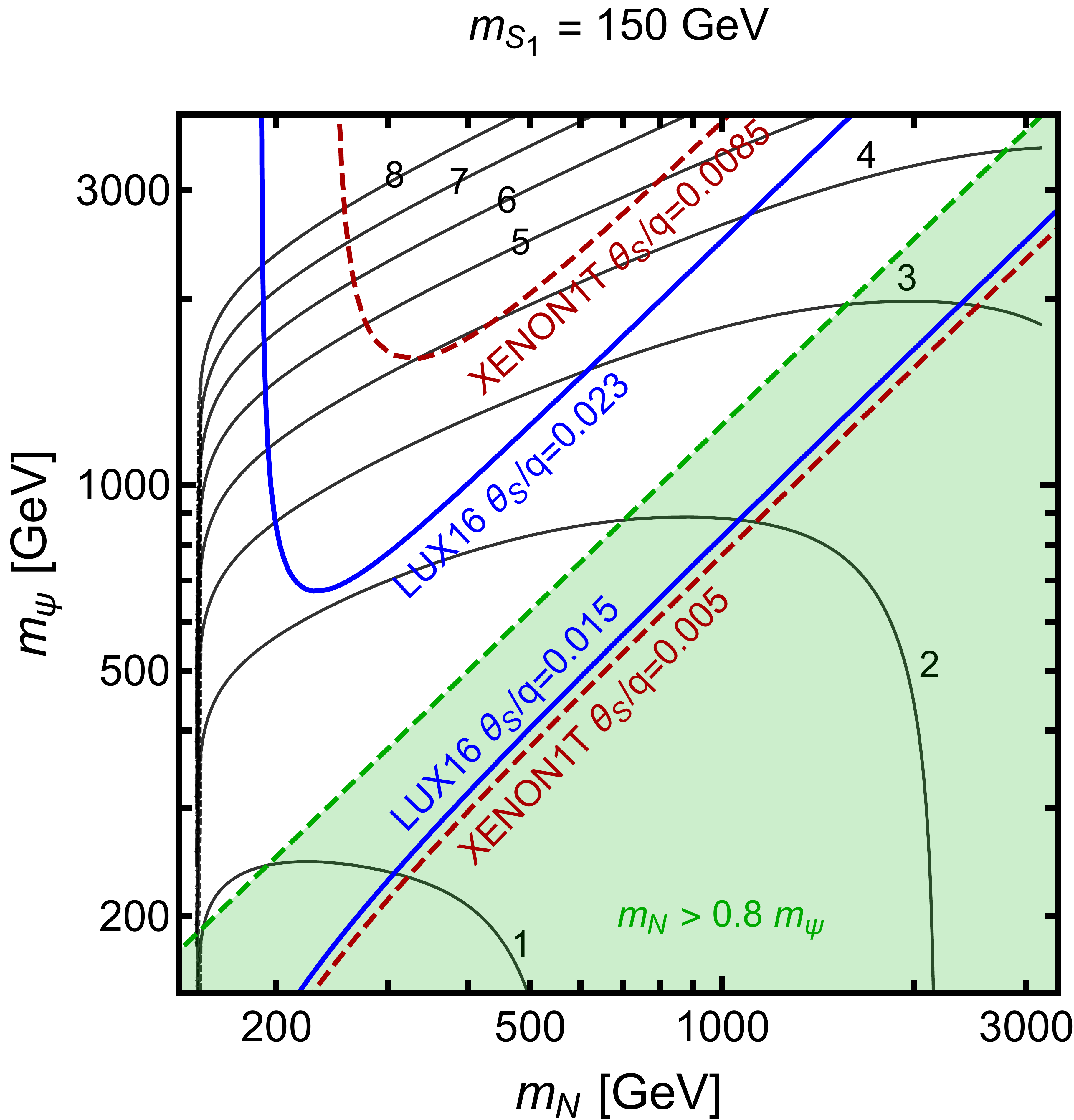}
\caption{The Yukawa coupling $y_S$ required to obtain the correct relic density of $N$. Existing bounds and future prospects for direct detection are also shown. These depend on $\theta_S$, the mixing angle between $S_1$ and the SM scalar boson $h$.\label{fig:NNss}}
\end{figure}

Let us first consider the case $m_{S_1} < m_N$. In this case, the dominant process for dark matter annihilation in the early Universe is:
\begin{equation*}
\includegraphics[width=2cm]{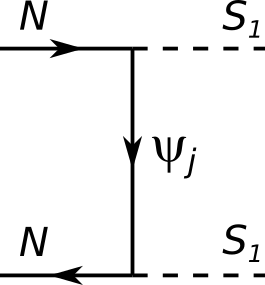}
\end{equation*}
Since the eigenstate $N$ has a large overlap with $R_0$ and the clockwork gears $\psi_i$ have O(1) overlap with $L_1$, the couplings involved in this diagram are O(1), generated from the Yukawa interaction~\footnote{In the limit $q \gg 1$ the analogous interactions involving $L_{>1}$ are instead subleading.} $y_S \, S_1 \overline{L}_1 R_0$. Thus, this process is in thermal equilibrium in the early Universe, freezing out when $N$ is non-relativistic. \emph{Although decaying dark matter, $N$ is a WIMP.} Since in the clockwork spirit all Lagrangian couplings are expected to be O(1), the WIMP miracle implies that $N$ has to be at the weak scale, up to the TeV range. This is shown in Fig.~\ref{fig:NNss}, where the Yukawa coupling $y_S$ that yields the correct relic density is plot. For instance, for $m_{S_1} = 150 \, \text{GeV}$, the perturbativity condition $y_S < \sqrt{4 \pi} \simeq 3.5$ requires $m_N, m_\psi \lessapprox \unit[2]{TeV}$.

\begin{figure}
\centering
\includegraphics[width=7cm]{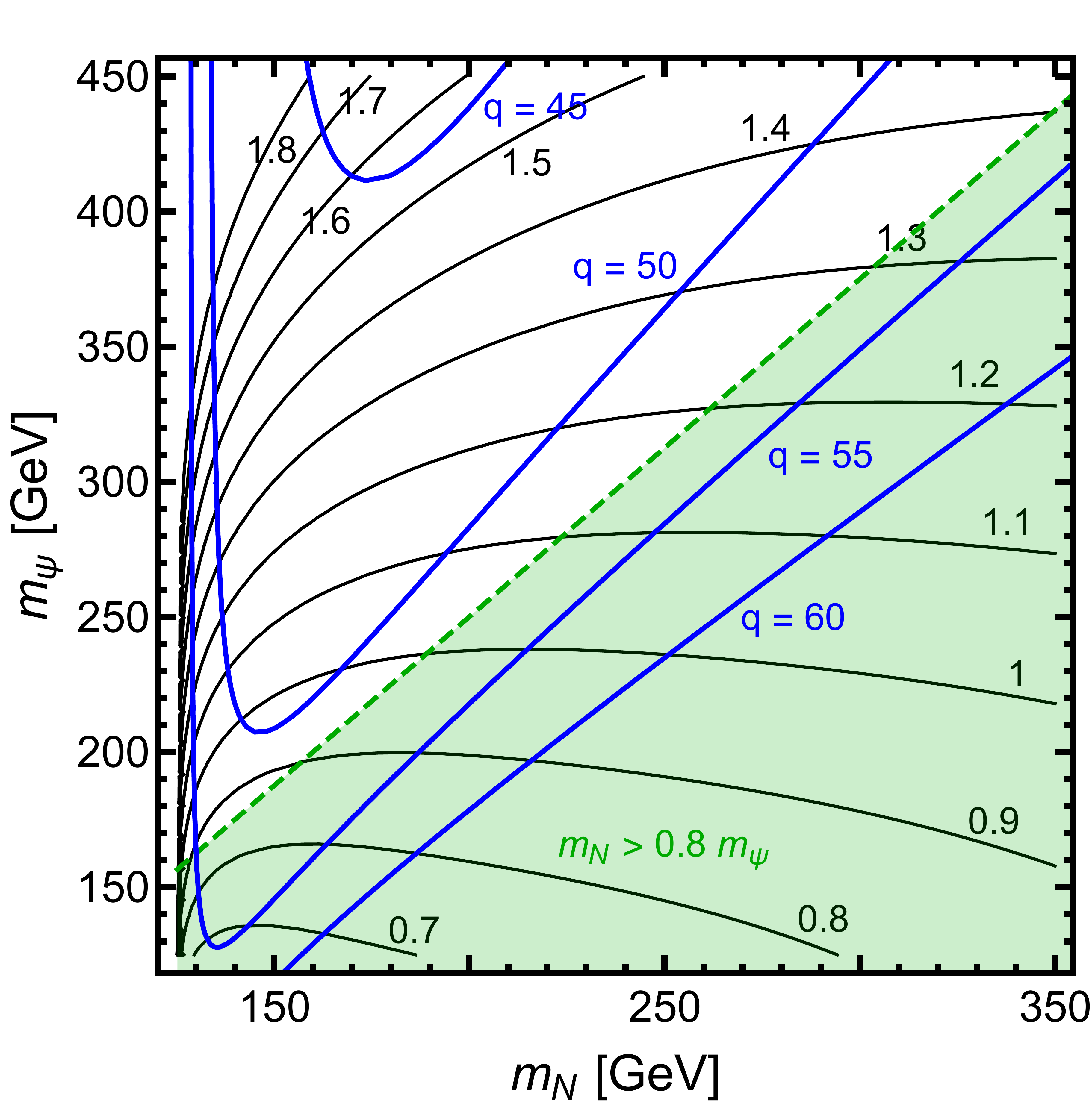}
\caption{The effective coupling $y_S \theta_S$ required to obtain the correct relic density of $N$. Existing bounds and future prospects for direct detection are also shown, depending on the value of $q$. \label{fig:NNhh}}
\end{figure}

In alternative, we may consider the case in which $m_S$ is so large that the $N N \rightarrow S_1 S_1, S_1 h$ are not kinematically allowed for non-relativistic $N$, so that the relevant process for dark-matter relic density is:
\begin{equation*}
\includegraphics[width=2cm]{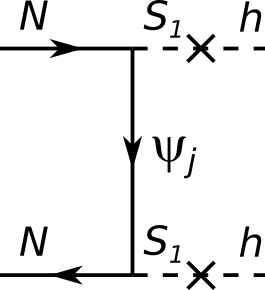}
\end{equation*}
The rate for this process is proportional to $(y_S \theta_S)^4$, where $\theta_S$ is the mixing of $S_1$ with the SM scalar $h$. The results for the effective coupling $y_S \theta_S$ yielding the correct relic density are shown in Fig.~\ref{fig:NNhh}. Since, as discussed below, the mixing of a single singlet scalar has to be less than $\approx 0.4$ from collider experiments, in the case of $N$ singlets with universal mixing one would have $\theta_S \lessapprox 0.4/\sqrt{N}$.  In this case the required values of $y_S$ from Fig.~\ref{fig:NNhh} would be non-perturbative. Therefore, this scenario works in the presence of non-universal mixing angles for the scalars, or near the $h$ or $S$ resonances, in this case also for universal mixing. The remaining intermediate scenario in which the dominant process is $N N \rightarrow S_1 h$ is discussed in Ref.~\cite{Hambye:2016qkf}. 

In addition to direct-detection experiments, the clockwork dark matter framework discussed here has a very rich phenomenology. First of all, peculiar signals at indirect-detection experiments are possible, because of the unstable nature of dark matter. For instance, the decay $N \to h \nu$ would give rise to a monochromatic neutrino line. An important feature of the clockwork construction that we are discussing is the presence of a band of fermionic states, the clockwork gears $\psi_i$, that \emph{need to be} in the TeV range or less. From the phenomenological point of view, these are essentially a collection of (pseudo-Dirac) heavy neutrinos in the observable mass range, with Yukawa couplings that are large, in the clockwork spirit. Therefore, the relevant indirect constraints are from electroweak precision tests (EWPT), $|B_{l\psi}|^2 \equiv y^2 v^2/(2 m_\psi^2) \lessapprox 10^{-3}$ and lepton flavour violation (LFV), $BR(\mu \to e \gamma) \ \approx \ 8 \times 10^{-4} \, |B_{e\Psi}|^2 \, |B_{\mu\Psi}|^2 < 4.2 \times 10^{-13}$, although the latter depend on the flavour structure of the couplings. LFV signatures are particularly promising, since planned experiments will improve the existing limits by several orders of magnitude in the near future. As for direct searches at colliders, lepton-number conserving processes will be tested, for couplings allowed by EWPT, up to $m_\psi \approx 200$ GeV with 300 $\mathrm{fb}^{-1}$ of data at LHC, whereas lepton-number violating processes will be tested up to $m_\psi \approx 300$ GeV.

\section{Clockwork from a flat extra dimension}
The structure of fields and couplings~\eqref{eq:clockwork_scheme} calls for some justification. Clearly, a possibility is that this is the discretized ``deconstructed'' version of an extra dimension, as suggested in~\cite{Giudice:2016yja}. There, in the fermionic case, the clockwork Lagrangian is obtained by discretizing a single massless fermion in 5D, with a metric $d s^2 = e^{4 k Z /3} (d x^2 + d Z^2)$, where $Z$ is the coordinate along the 5th dimension.

In Ref.~\cite{Hambye:2016qkf} we have shown that this is not the only possibility. We have considered a \emph{massive} fermion field in 5D, with bulk mass $M$, and a \emph{flat} metric, with $0 \leq Z \leq R$. In this case, the light (chiral) mode is introduced separately on one of the branes, say $Z=0$, and the SM leptons on the other one, sat $Z=R$. Equivalently, as in the subsequent construction of Ref.~\cite{Craig:2017cda}, the chiral mode can be seen as a boundary term for the bulk action. After discretizing the extra dimension in $N$ points with lattice spacing $a = R/N$, taking into account the appropriate Wilson term that avoids double counting, we find the 4D Lagrangian
\begin{equation}
\mathcal{L} \ \supset \ \sum_{i=0}^{N-1} \frac{1}{a} \,  \overline{L}_{i+1} R_{i} - \sum_{i=1}^{N} \bigg( \frac{1}{a} + M \bigg) \, \overline{L}_i R_i \;.
\end{equation}
We see that the clockwork setup~\eqref{eq:clockwork_scheme} is realized, by identifying
\begin{equation}
m \equiv \frac{1}{a} \;,\qquad  q m \equiv \frac{1}{a} + M \;.
\end{equation}
The continuum limit is obtained for $N \to \infty$, i.e.~$a \to 0$ with $R$ fixed. In this case $q \to 1$ and the total clockwork suppression remains finite
\begin{equation}
q^N = \left( 1 + \frac{\pi R M}{N} \right)^N \ \to \ e^{\pi R M}
\end{equation}
i.e.~a sensible continuum limit do indeed exist. In this framework, also the scalars $S_i$, $C_i$ in~\eqref{eq:scheme_scalars} can come from the discretization of a single field in the bulk. In particular, the fields $C_i$ could originate form a single 5D scalar with a Yukawa coupling with the fermion in the bulk, whereas the fields $S_i$ could come from the link variables of a 5D abelian gauge field, under which the bulk fermion is charged, and these from the 4D perspective are pseudo-scalars.

\section{Conclusions}
An unstable dark-matter candidate requires a huge suppression of its decay processes to comply with stability on cosmological time scales and bounds from indirect-detection experiments. I have discussed how the clockwork mechanism can provide that, without decoupling the relevant physics. In particular, this framework \emph{requires}~\footnote{not ``can accommodate''.} the presence of new clockwork states at the TeV range or less and large couplings with the Standard Model, in order to obtain the correct relic density. The very same states mediate both DM annihilation in the early Universe, which fixes the relic density, and the clockwork-suppressed decay processes, thus providing a rather unique connection between them. From the phenomenological point of view, this construction yields rich signatures in a number of experiments. The key aspect is the presence of a band of pseudo-Dirac right-handed neutrinos in the TeV range or less, that can be searched for at colliders and lepton-flavour-violation experiment.

I have also discussed how this setup could emerge from the deconstruction of a flat extra dimension. This (de)construction differs form the original one of Ref.~\cite{Giudice:2016yja}, which instead makes use of a curved metric. 
After we originally presented this in Ref.~\cite{Hambye:2016qkf}, it has been shown very recently~\cite{Giudice:2017suc} that the two constructions are indeed equivalent, at least in the bosonic case.

Finally, I conclude mentioning that Majorana neutrino masses can be incorporated in the same framework discussed here, and that these are clockwork-suppressed too. For more details on this and the rest of the material discussed here, I refer the reader to the original work~\cite{Hambye:2016qkf}, where these matters were first discussed.

\section*{Acknowledgements}
I thank the organizers of the Moriond Electroweak 2017 conference, and in particular Jean-Marie Fr\`ere, for providing a beautiful environment to discuss fundamental physics. I also thank Thomas Hambye and Michel Tytgat for collaboration on the work discussed in this note. My work has been supported by a postdoctoral fellowship of ULB and by the Belgian Federal Science Policy IAP P7/37.

\end{document}